\documentstyle[amssymb,prl,aps]{revtex}

\draft

\begin{document}
\title{''Averaged'' statistical thermodynamics, energy equipartition and the third
law}
\author{Vesselin I. Dimitrov\thanks{%
E-mail: vesko@phys.uni-sofia.bg}}
\address{Faculty of Physics, Sofia University, BG-1164 Sofia, Bulgaria}
\date{\today }
\maketitle

\begin{abstract}
Arguments are presented that the assumption, implicit to traditional
statistical thermodynamics, that at zero temperature all erratic motions
cease, should be dispensed with. Assuming instead a random ultrarelativistic
unobservable motion, similar to zitterbewegung, it{\ is demonstrated that in
an ideal gas of classical particles the energy equipartition fails in a way
that complies with the third law of thermodynamics.}
\end{abstract}


\subsection*{Introduction}

Velocity cannot be measured. Readily measurable are positions and time
intervals. Textbook definitions of velocity involve a limiting procedure in
the definition of derivatives. Mathematically, such a procedure makes
perfect sense; its operational meaning, however, is obscure. To clarify this
statement, let me quote from Einstein's note\cite{Einstein 1907} on brownian
motion (in a loose translation):

''{\it ..Since an observer can never be aware of pieces of path passed in
arbitrarily small intervals of time (regardless of the methods and means
at his disposal), he will always accept as a momentary velocity some average
velocity. It is clear, though, that the velocity determined in this way does
not correspond to any objective property of the considered motion..}''

Classical mechanics and, indirectly, classical statistical thermodynamics
rely heavily on the notion of a Lagrangian. Being defined as a function of
particle's coordinates and velocities, the Lagrangian cannot be measured,
either. This circumstance by no means prevents its use in establishing the
classical formalism; it is rather the interpretation of observations in
terms of the theory where difficulties may be encountered.

One doesn't have to look far for examples of such difficulties. About the
beggining of our century, one of the symptoms of the ''crisis in physics''
was that observations of the properties of thermal radiation were in
conflict with energy equipartition -- a rigorous result of classical
statistical thermodynamics. To anyone sharing the ideas of Poincar\'{e}
about the meaning of a physical theory\cite{Poincare}, this could only mean
that theory has been inconsistently applied to the interpretation of data;
the physical community at that time, however, took the position that
classical theory was wrong and needed modification, and eventually, quantum
mechanics was born. It is not my purpose here to discuss the merits and
shortcomings of quantum theory; in what follows I will rather try to explore
the relevance of possible motions beyond the detectability limits of one's
equipment to the issue of energy equipartition entirely within classical
statistical thermodynamics.

The idea that there may exist motion beyond what can be readilly observed is
by no means new. Shortly after the discovery of Dirac's equation,
Schr\"{o}dinger pointed out that it predicts a curious phenomenon --
ultrarelativistic erratic motion of the center of charge about the center of
mass, that has been called {\it zitterbewegung}\cite{Schroedinger}. Since
then, many aspects of both the quantum and classical versions of
zitterbewegung have been studied (see e.g. \cite{zbw}). However, to the
knowledge of the present author, the implications of this irregular motion
for thermodynamics have not received much attention so far. Hence, one way
to define the purpose of the present work is studying the relevance of
classical zitterbewegung for the issue of energy equipartition in classical
statistical thermodynamics.

It should be noted of course, that equipartition of energy fails even in
traditional statistical thermodynamics as soon as the motion becomes
relativistic, and it does so at two counts. The relativistic relation
between energy and velocity of a point particle does not posses the
additivity property with respect to its three spatial degrees of freedom.
Furthermore, the relativistic relation between energy and temperature
involves particle's rest mass explicitly\cite{rthermo}, hence in a
relativistic gas at certain temperature, particles of different rest mass
have different average energy. Therefore, it will be the purpose of the
present work to explore the issue of energy equipartition in a situation
where the apparent velocity is small enough compared to velocity of light,
such as to suggest nonrelativistic treatment.

\subsection*{Kinematics: Observable and hidden velocity}

It should be clearly understood that the problem with the equipment
mentioned in the introduction, is not merely a technical one but is
connected with fundamental limitations of the theory. Indeed, it is well
known that classical electrodynamics of an electron becomes
self--contradictory whenever distances smaller than classical electron
radius $r_e$ and time scales smaler than $\tau _e=2r_e/3c$ are involved\cite
{Jackson}. Thus a prudent approach would separate electron's velocity ${\bf %
\beta }$ into a ''mean'' velocity

\begin{equation}
{\bbox{\beta }}_{T}(t)=\frac{1}{\tau _{e}}\int_{0}^{\tau _{e}}dt^{\prime }%
\bbox{\beta }(t-t^{\prime })=\frac{1}{c\tau _{e}}\left[ {\bf r}(t)-{\bf r}%
(t-\tau _{e})\right]  \label{eq1}
\end{equation}
and the rest. Since we have no {\it apriori} knowledge about the actual
velocity of the electron, we safely assume a relativistic situation.
Relativistic kinematics imply
\begin{equation}
\bbox{\beta }_{\Vert }=\frac{\bbox{\beta }_{T}+\bbox{\beta }_{0\Vert }} {1+%
\bbox{\beta }_{0}\cdot \bbox{\beta }_{T}}\;\;\;\bbox{\beta }_{\bot }= \frac{%
\bbox{\beta }_{0\bot }\sqrt{1-\beta _{T}^{2}}}{1+\bbox{\beta }_{0}\cdot %
\bbox{\beta }_{T}}  \label{eq2}
\end{equation}
where $\bbox{\beta }_{0}$ is the ''unobservable'' velocity and the $\Vert $
and $\bot $ components are with respect to $\bbox{\beta }_{T}$. From this
expression one finds for the momentum

\begin{eqnarray}
{\bf p} &=&m_0c\frac{\bbox{\beta }}{\sqrt{1-\beta ^2}}=  \nonumber \\
&=&\frac{m_0c}{\sqrt{1-\beta _0^2}}\frac{\bbox{\beta }_T}{\sqrt{1-\beta _T^2}%
}+\frac{m_0c}{\sqrt{1-\beta _T^2}}\frac{\bbox{\beta }_{0\Vert }+\bbox{\beta }%
_{0\bot }\sqrt{1-\beta _T^2}}{\sqrt{1-\beta _0^2}}  \label{eq3}
\end{eqnarray}
and for the kinetic energy

\begin{equation}
E_k=m_0c^2\left( \frac 1{\sqrt{1-\beta ^2}}-1\right) =m_0c^2\left( \frac{1+%
\bbox{\beta }_0\cdot \bbox{\beta }_T}{\sqrt{1-\beta _0^2}\sqrt{1-\beta _T^2}}%
-1\right)  \label{eq4}
\end{equation}
Lacking any knowledge of $\bbox{\beta }_0$ other than $\beta _0\leq 1$, one
can regard it as a random variable distributed isotropically. By virtue of
Eqs.(\ref{eq3}-\ref{eq4}) momentum and energy become random variables, too.
With this assumption the average momentum and the average energy turn out to
be

\begin{eqnarray}
&<&{\bf p}>\,=\,mc\frac{\bbox{\beta }_T}{\sqrt{1-\beta _T^2}}  \nonumber \\
&<&E_k>\,=\,mc^2\left( \frac 1{\sqrt{1-\beta _T^2}}-1\right)  \label{eq5}
\end{eqnarray}
where the average mass is given by

\begin{equation}
m=\,<\frac{m_0}{\sqrt{1-\beta _0^2}}>  \label{eq6}
\end{equation}

Apparently, the average momentum and energy have the cusomary dependence on
the mean velocity. The only effect of the unobservable velocity on them is
that the particle's mass appears to be larger. If observer's equipment can
only measure average momenta and energy, no experiment can reveal either the
presence of $\bbox{\beta }_0$ or the value of $m_0$. This situation
radically changes as soon as one considers fluctuations.

\subsection*{Fluctuations of energy and apparent temperature}

Within statistical thermodynamics, the temperature $\theta $ can be related
to the dependence of the energy fluctuations on the average energy \cite
{thermo}:

\begin{equation}
-\frac{\partial \overline{E}}{\partial \theta ^{-1}}=\,\overline{E^2}-%
\overline{E}^2  \label{eq7}
\end{equation}
where the overbar indicates thermal averaging. One way to look at this
relation is as a statistical definition of temperature, consistent with the
first and second principles of thermodynamics. For example, if energy is
distributed with probability density involving{\em \ only one parameter}
with dimensions of energy $F(E/E_0)$ (the cannonical distribution obviously
belongs to this class):

\begin{eqnarray*}
&&\overline{E}=\frac{\int dEEF(E/E_0)}{\int dEF(E/E_0)}=E_0\frac{\int dxxF(x)%
}{\int dxF(x)}\equiv aE_0 \\
&&\overline{E^2}=\frac{\int dEE^2F(E/E_0)}{\int dEF(E/E_0)}=E_0^2\frac{\int
dxx^2F(x)}{\int dxF(x)}\equiv bE_0^2 \\
&&\overline{E^2}-\overline{E}^2=(\frac b{a^2}-1)\overline{E}^2\equiv d%
\overline{E}^2
\end{eqnarray*}
where $a,b$ and $d$ are non-negative dimensionless constants depending on
the particular form of the distribution function. Plugging the result of the
last row above into Eq.(\ref{eq7}) and solving with boundary condition $%
\lim_{\theta \rightarrow \infty }\overline{E}=\infty $, we obtain the
familiar relation

\begin{equation}
\overline{E}=\frac \theta d  \label{eq8}
\end{equation}
For the canonical distribution $d$ is, of course, equal to $2/N$ where $N$
is the number of degrees of freedom, and Eq.(\ref{eq8}) expresses the energy
equipartition property of classical statistical thermodynamics.

The relevance of the above definition of temperature to our discussion
becomes apparent when we allow $\bbox{\beta }_T$ above to be a subject of
thermal fluctuations, statistically independent from the fluctuations of $%
\bbox{\beta }_0$. In other words, we shall be interested in exploring the
effects of the unobservable motion on the statistical properties of the
observable one. Straightforward manipulation using Eq.(\ref{eq4}) and the
isotropy of $\bbox{\beta }_0$ yields, with $\gamma _T\equiv (1-\beta
_T^2)^{-1/2}$

\[
<E_k^2>\,=\frac{4<m^2>c^4-m_0^2c^4}3\left( \gamma _T-1\right) ^2+2\frac{%
<m^2>c^4-m_0^2c^4}3\left( \gamma _T-1\right)
\]
Combining this with Eq.(\ref{eq5}) and performing thermal averaging, we
arrive at

\begin{eqnarray}
&<&\overline{E_k^2}>-<\overline{E_k}>^2=  \nonumber \\
&=&\frac{4<m^2>-m_0^2}{3m^2}\left( mc^2\right) ^2\overline{\left( \gamma
_T-1\right) ^2}-<mc^2>^2\overline{\left( \gamma _T-1\right) }^2+  \label{eq9}
\\
&&+2\frac{<m^2>-m_0^2}{3m^2}\,mc^2\left[ mc^2\overline{\left( \gamma
_T-1\right) }\right]  \nonumber
\end{eqnarray}

Eq.(\ref{eq9}) reveals two distinct effects of the presence of the
unobservable velocity $\bbox{\beta }_0$. First, the coefficient $\frac{%
4<m^2>-m_0^2}{3m^2}$ appears whereas in the absence of $\bbox{\beta }_0$ one
would have a unit factor instead; and second, the whole last term in the
right--hand side of Eq.(\ref{eq9}), is absent in the traditional approach.

At this point it is convenient to make two further assumptions concerning
the statistics of the unobservable velocity. First, given the general
validity of statistical thermodynamics, we can safely assume that $\beta _0$
is distributed according to the relativistic Maxwell law with some
temperature $\theta _0$. Simple calculation then produces

\begin{eqnarray*}
&&m\equiv \,<m>\,=m_0\left[ \frac 3\alpha +\frac{K_1(\alpha )}{K_2(\alpha )}%
\right] \\
&<&m^2>\,=m_0^2\left[ \frac{12}{\alpha ^2}+1+\frac{3K_1(\alpha )}{\alpha
K_2(\alpha )}\right]
\end{eqnarray*}
where $K_n$ is a modified Bessel function of the second kind and $\alpha
\equiv m_0c^2/\theta _0$. Second, let us take $\theta _0$ large enough such
as to have $\alpha \ll 1$. This means that the bare mass $m_0$ is assumed
small compared to the apparent mass $<m>$. For the time being let us
consider this second assumption simply as a means for simplifying the
resulting expressions. Thus, we calculate

\[
\frac{<m^2>}{m^2}=\frac 43+O(\alpha ^2)
\]
and rewrite Eq.(\ref{eq9}) in lowest (zeroth) order of $\alpha $ as$\ $%
\begin{eqnarray}
&<&\overline{E_k^2}>-<\overline{E_k}>^2=  \nonumber \\
&=&\left[ (\frac 43)^2(1+\varkappa )-1\right] <\overline{E_k}>^2+\frac 23%
\times \frac 43<mc^2><\overline{E_k}>  \nonumber \\
&&\varkappa \equiv \frac{\overline{\left( \gamma _T-1\right) ^2}}{\overline{%
\left( \gamma _T-1\right) }^2}-1\in [0,1]  \label{eq10}
\end{eqnarray}

Using this in Eq.(\ref{eq7}), we can ascribe an apparent temperature
associated with $<E_k>$, i.e. with the observable velocity $\beta _T$.
Requiring, as above, $\lim_{\theta _T\rightarrow \infty }<\overline{E_k}%
>\,=\infty $ , solving the differential equation, and defining $\varepsilon
_0=4/3\,<mc^2>$, we easily obtain

\begin{equation}
<\overline{E_k}>\,=\frac{2/3}{(\frac 43)^2(1+\varkappa )-1}\,\frac{%
\varepsilon _0}{\exp \left( \frac{\varepsilon _0}{3\theta _T/2}\right) -1}%
\equiv \frac 1{\widetilde{\varkappa }}\,\frac{\varepsilon _0}{\exp \left(
\frac{\varepsilon _0}{3\theta _T/2}\right) -1}  \label{eq11}
\end{equation}

Now, this is a very curious result that needs few comments. The first one
concerns our definition of $\varepsilon _0$. The factor $4/3$, appearing in
connection with electron mass, has aquired certain notoriety in the context
of a discussion about the electromagnetic contribution to the mass\cite
{emass}; its origin however is most clearly explained in relativistic
thermodynamics: in relativistic systems the quantity that, together with the
three components of momentum, forms a 4--vector, is the enthalpy rather than
the energy\cite{Kibble}. Now it is easy matter to verify that, in zeroth
order of $\alpha $, the enthalpy per particle is exactly $4/3$ times the
energy per particle\cite{rthermo}, coinciding with our $\varepsilon _0$ from
above.

The second comment concerns the behaviour of an ideal gas of classical
particles with $\beta _0\neq 0$. It is well known that traditional ideal
gas, where $\beta _0$ is absent, is out of touch with the third law of
thermodynamics; actually it is true for every system satisfying the energy
equipartition theorem. Historically, energy equipartition failure has been
first demonstrated for blackbody radiation, and has been explained by
introducing the notion of quanta. Thus, the fact that experimentally
observed thermodynamic systems obey the third law of thermodynamics and
violate energy equipatrition theorem is usually regarded as a proof of the
quantum nature of those systems. In a marked contrast to this, our Eq.(\ref
{eq11}) implies the following expression for the specific heat capacity of
an ideal gas of classical particles with $\beta _0\neq 0$:

\begin{equation}
c_V\equiv \frac{\partial <\overline{E_k}>}{\partial \theta _T}=\frac 2{3%
\widetilde{\varkappa }}\frac{\varepsilon _0^2}{\theta _T^2}\frac{\exp \left(
\frac{\varepsilon _0}{3\theta _T/2}\right) }{\left[ \exp \left( \frac{%
\varepsilon _0}{3\theta _T/2}\right) -1\right] ^2}  \label{eq12}
\end{equation}
whereas for traditional ideal gas $c_V=3/2$ holds. Letting the apparent
temperature $\theta _T$ go to zero, we observe that $c_V$ defined by Eq.(\ref
{eq12}) goes to zero, too. Now this behaviour is just the crucial aspect of
the third law\cite{Boyer0}, hence the presence of the $\beta _0$ causes a
failure of energy equipartition in a way that makes our classical system
compliant with the third law of thermodynamics.

As a last comment, a remark concerning the high--temperature behaviour of
Eqs.(\ref{eq11},\ref{eq12}) is in order. For $\theta _T\rightarrow \infty $
we have

\[
<\overline{E_k}>\simeq \frac 1{2\widetilde{\varkappa }}\,3\theta
_T\;\;c_V\simeq \frac 3{2\widetilde{\varkappa }}
\]
This is about as far as we can go without further information about the
distribution of $\beta _T$. Assuming for a moment relativistic Maxwell
distribution for $\beta _T$, we find $\varkappa =1/3$ and $\widetilde{%
\varkappa }\simeq 2.06$. Thus, at high temperature an ideal gas of particles
with ultrarelativistic $\beta _0$ would appear, due to the factor $%
\widetilde{\varkappa }^{-1}$, like traditional ideal gas with about half of
its degrees of freedom ''frozen'' and not contributing to its heat capacity.
This, as well as the form of Eqs.(\ref{eq11},\ref{eq12}) is curiously
reminiscent of the properties of an ideal Bose--Einstein gas.

\subsection*{Discussion}

A meaningful physical theory should posses certain ''stability'' of its
predictions with respect to assumptions that cannot be experimentally tested%
\cite{Popper}. One such assumption in classical statistical thermodynamics
is that the apparent velocity of a particle is identical to its actual
velocity. Whether this is true or not is an issue that cannot be settled
down within classical theory, due to the limitations to scales larger than
classical electron radius and corresponding time intervals, inherent to
classical electrodynamics. In traditional classical physics this assumption,
in a form stating that at zero temperature all erratic motion ceases, is
always implicitly done. In the case of thermodynamics of classical
electromagnetic radiation and charged harmonic oscillators, it has been
shown that dispensing with it not only doesn't inflict basic thermodynamic
principles, but even makes possible an entirely classical understanding of
the phenomenology of blackbody radiation\cite{Cole} In a more general
context, admitting the existence of a Lorentz--invariant classical random
electromagnetic field at zero temperature results in an interesting
classical theory, called ''Stochastic Electrodynamics'', which has proven
capable of dealing with a number of phenomena, usually believed to belong to
the quantum realm, but still lacks the generality of quantum theory and
possesses problems of its own (for a review with extensive references, see
\cite{sed}). This, combined with the results of the present work, clearly
demonstrates the instability of the predictions of traditional classical
thermodynamics with respect to the assumption of the absense of erratic
motion at zero temperature. It, therefore, should be purged from the theory,
and it is our conjecture that, without it, classical statistical
thermodynamics would provide comprehensible description of all phenomena
currently believed to require quantum treatment. Whether this is true, and
how this could be done, remains to be elaborated on.

\subsection*{Acknowledgement}

The author wishes to thank the foundation ''Bulgarian Science and Culture''
for its support.

\end{document}